\documentclass[10pt,onecolumn,a4paper]{IEEEtran}
\usepackage[latin1]{inputenc}
\usepackage{amssymb}
\usepackage{latexsym}
\usepackage{amsfonts}
\usepackage{amsbsy}
\usepackage{amsmath}
\usepackage{times}
\usepackage{epsfig,epsf,times,amssymb,amsmath}
\usepackage{color}

\input epsf
\title{Low-Complexity Channel Estimation with Set-Membership Algorithms for Cooperative Wireless Sensor Networks}
\author{Tong Wang, Rodrigo C. de Lamare, and Paul D. Mitchell}

\begin{document}
\maketitle
\thispagestyle{empty}
\begin{abstract}

In this paper, we consider a general cooperative wireless sensor network (WSN) with multiple hops and the problem of channel estimation. Two matrix-based set-membership algorithms are developed for the estimation of the complex matrix channel parameters. The main goal is to reduce the computational complexity significantly as compared with existing channel estimators and extend the lifetime of the WSN by reducing its power consumption. The first proposed algorithm is the set-membership normalized least mean squares (SM-NLMS) algorithm. The second is the set-membership recursive least squares (RLS) algorithm called BEACON. Then, we present and incorporate an error bound function into the two channel estimation methods which can adjust the error bound automatically with the update of the channel estimates. Steady-state analysis in the output mean-squared error (MSE) are presented and closed-form formulae for the excess MSE and the probability of update in each recursion are provided. Computer simulations show good performance of our proposed algorithms in terms of convergence speed, steady state mean square error and bit error rate (BER) and demonstrate reduced complexity and robustness against the time-varying environments and different signal-to-noise ratio (SNR) values.\\
\\

\end{abstract}

\section{Introduction}
Recently, there has been a growing research interest in wireless sensor networks (WSNs) because their unique features allow a wide range of applications in the areas of military, environment, health and home \cite{Akyildiz}. They are usually composed of a large number of densely deployed sensing devices which can transmit their data to the desired user through multihop relays \cite{Laneman}. Low complexity and high energy-efficiency are the most important design characteristics of communication protocols \cite{Mitchell} and physical layer techniques employed for WSNs. The performance and capacity of WSNs can be significantly enhanced through exploitation of spatial diversity with cooperation between the nodes \cite{Laneman}. In a cooperative WSN, nodes relay signals to each other in order to propagate redundant copies of the same signals to the destination nodes. Among the existing relaying schemes, the amplify-and-forward (AF) and the decode-and-forward (DF) are the most popular approaches \cite{Hong}. Due to limitations in sensor node power, computational capacity and memory \cite{Akyildiz}, some power-constrained relay strategies \cite{Khajehnouri,Krishna} and power allocation methods \cite{Li} have been proposed for WSNs to obtain the best possible SNR or best possible quality of service (QoS) at the destinations. Most of these ideas are based on the assumption of perfect synchronization and available channel state information (CSI) at each node \cite{Akyildiz}.
Therefore, more accurate estimates of the CSI will bring about better performance in WSNs.

The normalized least mean squares (NLMS) estimation method is appropriate for WSNs due to its simplicity. However, the main problem of the NLMS is that the tradeoff between convergence speed and steady state performance is achieved through the introduction of a step size \cite{Haykin}. It is not possible to achieve the best solution on these two aspects using a conventional NLMS estimation method. Channel estimation with the NLMS algorithm can be improved by introducing the set-membership filtering (SMF) framework \cite{Lamare} which modifies the objective function of the NLMS algorithm. It specifies an error bound on the magnitude of the estimation error, which can make the step size adaptive. Therefore the SM-NLMS channel estimation method can achieve good convergence and tracking performance for each update. A SM-NLMS channel estimation algorithm for cooperative WSNs is proposed in \cite{Wang}. Compared with the NLMS channel estimation method, the RLS
channel estimator can provide better performance in terms of the
convergence speed and steady state \cite{Haykin}. However, it is
not suitable for WSNs due to its high computational complexity
\cite{Haykin}. In order to overcome this shortcoming, the SMF framework can be also
introduced to devise a computationally efficient version of the
conventional RLS channel estimation method, called BEACON channel
estimation. It can be considered as a constrained optimization
problem where the objective function is the least squares (LS)
cost function and the constraint is a bound on the magnitude of
the estimation error. As a result, an adaptive forgetting factor
can be derived to achieve the optimal performance for each update.
Most importantly, the set-membership (SM) algorithms possess a feature that allows
updating for only a small fraction of the time, expressed as the update rate (UR). Therefore, the UR of the two SM channel estimation algorithms decreases due to the data-selective update which can
reduce the computational complexity significantly and extend the
lifetime of the WSN by reducing its power consumption.

The biggest issue for the SM channel estimation is the appropriate selection of the error bound, because it has a critical effect on the estimation performance. For SM-NLMS channel estimation, the extreme settings of the bound, namely, overbounding (the error bound being too large) and underbounding (the error bound being too small) will result in performance degradation \cite{Guo,Galdino}. In practice, the bound depends on the environmental parameters such as the SNR. It is very difficult to determine the optimal error bound accurately because there is usually insufficient knowledge about the underlying system. For the BEACON channel estimation, the value of the error bound can be varied to trade off achievable performance against computational complexity
\cite{Nagara}. A higher error bound would result in lower UR but worse performance. For WSNs the aim is to achieve an acceptable CSI quickly with low power consumption. Therefore, the bound for BEACON channel estimation should be adjusted to ensure good estimation performance, lower computational complexity and a low UR. Also, the required error bound may be time variant due to changing environmental conditions.

In this paper, we develop two matrix-based SM algorithms for channel
estimation in cooperative WSNs using the AF cooperation protocol.
The major novelty in these algorithms presented here is that they
are matrix-based SM channel estimation algorithms as opposed to
vector-based SM techniques for filtering applications
\cite{Gollamudi1,Nagara,Dasgupta}. Therefore we specify a bound on
the norm of the estimation error vector instead of the magnitude of
the scalar estimation error. Then, a novel error bound function is
introduced to change the error bound automatically in order to
obtain optimal performance with the proposed SM channel estimation.
Furthermore, we propose analytical expressions of the steady-state
output excess mean-square error (MSE) of the two SM channel
estimation methods. Further novelty in this analysis is that we
employ the chi-square distribution to describe the probability of
the update for estimating the channel matrix as opposed to the
Gaussian distribution for estimating the filter vector
\cite{Paulo,Markus}. A key contribution of this paper is the
consideration of techniques to  reduce the complexity of the channel
estimation for WSNs.

This paper is organized as follows.  Section II describes the general cooperative WSN system model and its constrained form. Section III introduces two conventional channel estimation methods for reference. Section IV proposes two channel estimation methods using the SMF framework and presents an error bound function which tunes the error bound automatically. Section V contains the analysis of the steady-state output excess MSE and the computational complexity. Section VI presents and discusses the simulation results, while Section VII provides some concluding remarks.

\section{Cooperative WSN System Model}

Consider a general m-hop wireless sensor network (WSN) with
multiple parallel relay nodes for each hop, as shown in Fig.
1.  The WSN consists of $N_s$ sources, $N_d$ destinations and
$N_r$ relays which are separated into $m-1$ groups:
$N_{r(1)}$,$N_{r(2)}$, ... ,$N_{r(m-1)}$. All these nodes are
assumed to be within communication range. We will concentrate
on a time division scheme with perfect synchronization, for which
all signals are transmitted and received in separate time slots.
The sources first broadcast the $N_s\times1$ signal vector
\textbf{s} to the destinations and all groups of relays. We
consider an amplify-and-forward (AF) cooperation protocol in this
paper. Each group of relays receives the signal from the sources
and previous groups of relays, amplifies and rebroadcasts them to
the next groups of relays and the destinations. In practice, we
need to consider the constraints on the transmission policy. For
example, each transmitting node would transmit during only one phase. In our
WSN system, we assume that each group of relays transmits the
signal to the nearest group of relays and the destinations
directly. We can use a block diagram to indicate the cooperative
WSN system with these transmission constraints as shown in Fig. 2.

\begin{figure}[!htb]
\begin{center}
\hspace*{0em}{\includegraphics[width=12cm, height=9cm]{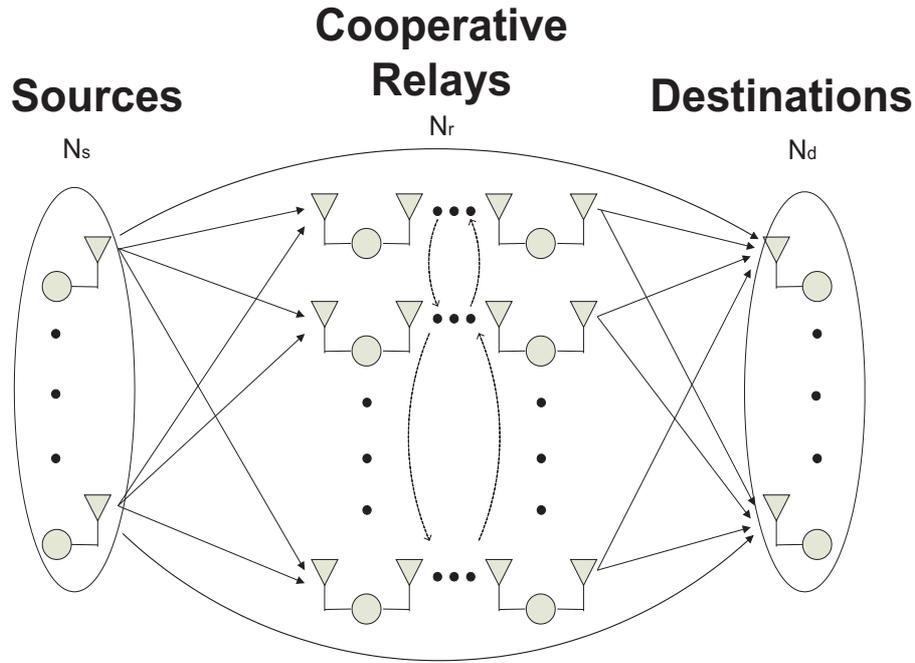}}
\vspace*{0.0em} \caption{An $m$-hop cooperative WSN with $N_s$
sources, $N_d$ destinations and $N_r$ relays.}
\end{center}
\end{figure}

\begin{figure}[!htb]
\begin{center}
\hspace*{0em}{\includegraphics[width=12cm, height=9cm]{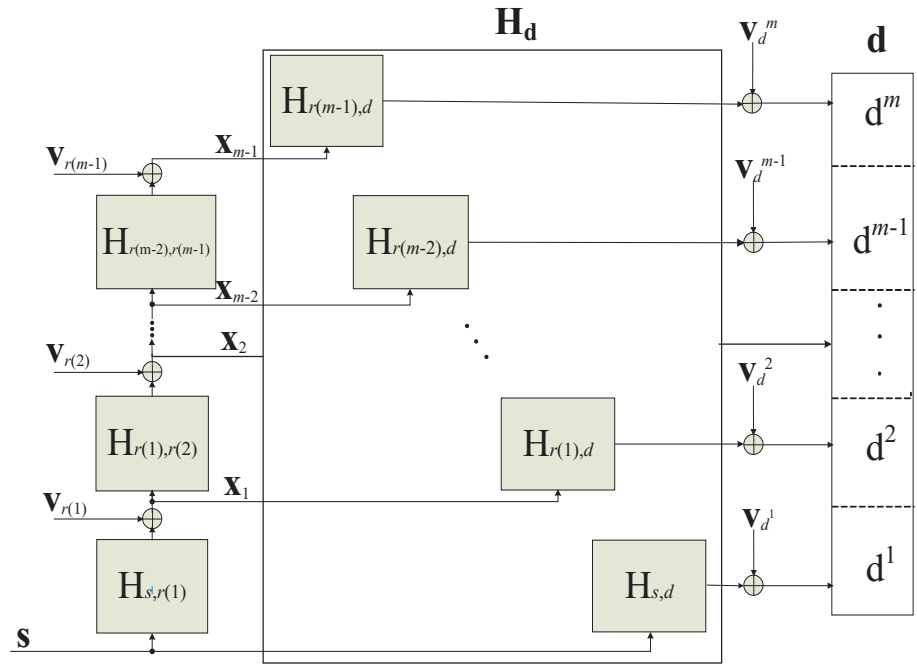}}
\vspace*{0.0em} \caption{Block diagram of the cooperative WSN
system with transmission constraints.}
\end{center}
\end{figure}

Let $\textbf{H}_{s,r(i)}$ denotes the  $N_{r(i)}\times{N_s}$
channel matrix between the sources and the $i$th group of relays,
$\textbf{H}_{r(i),d}$ denotes the $N_d\times{N_{r(i)}}$ channel
matrix between the $i$th group of relays and destinations, and
$\textbf{H}_{r(i-1),r(i)}$ denotes the
$N_{r(i)}\times{N_{r(i-1)}}$ channel matrix between two groups of
relays. The received
signal at the $i$th group of relays ($\textbf{x}_i$) and
destinations (\textbf{d}) for each phase can be expressed as:

Phase 1:\\
\begin{equation}
\textbf{x}_1=\textbf{H}_{s,r(1)}\textbf{s}+\textbf{v}_{r(1)}
\end{equation}
\begin{equation}
\textbf{d}^1=\textbf{H}_{s,d}\textbf{s}+\textbf{v}^1_d
\end{equation}

Phase 2:
\begin{equation}
\textbf{x}_2=\textbf{H}_{r(1),r(2)}\textbf{A}_1\textbf{x}_1+\textbf{v}_{r(2)}
\end{equation}
\begin{equation}
\textbf{d}^2=\textbf{H}_{r(1),d}\textbf{A}_1\textbf{x}_1+\textbf{v}^2_d
\end{equation}
\vdots

Phase $i$: ($i=2, 3, ... , m-1$)
\begin{equation}
\textbf{x}_i=\textbf{H}_{r(i-1),r(i)}\textbf{A}_{i-1}\textbf{x}_{i-1}+\textbf{v}_{r(i)}
\end{equation}
\begin{equation}
\textbf{d}^i=\textbf{H}_{r(i-1),d}\textbf{A}_{i-1}\textbf{x}_{i-1}+\textbf{v}^i_d
\end{equation}
\vdots

Phase $m$:
\begin{equation}
\textbf{d}^m=\textbf{H}_{r(m-1),d}\textbf{A}_{m-1}\textbf{x}_{m-1}+\textbf{v}^m_d
\end{equation}
where \textbf{v} is a zero-mean circularly symmetric complex
additive white Gaussian noise (AWGN) vector with covariance matrix
$\sigma^2\textbf{I}$. $\textbf{A}_i$ is a diagonal matrix whose
elements represent the amplification coefficient of each relay of
the $i$th group. The vectors $\textbf{d}^i$ and $\textbf{v}_d^i$
denote the received signal and noise at the destination nodes during
the $i$th phase, respectively. At the destination nodes, the received
signal can be expressed as:
\begin{equation}
\textbf{d}=\textbf{H}_d\textbf{A}\textbf{y}+\textbf{v}_d
\end{equation}
where,
\begin{equation}
 \textbf{d}=
 \begin{bmatrix}
 \textbf{d}^m\\
---\\
 \textbf{d}^{m-1}\\
---\\
 \vdots\\
---\\
 \textbf{d}^2\\
---\\
 \textbf{d}^1\\
 \end{bmatrix},~~~~~~
 \textbf{v}_d=
 \begin{bmatrix}
 \textbf{v}_d^m\\
---\\
 \textbf{v}_d^{m-1}\\
---\\
 \vdots\\
---\\
 \textbf{v}_d^2\\
---\\
 \textbf{v}_d^1\\
 \end{bmatrix},~~~~~~
 \textbf{y}=
 \begin{bmatrix}
 \textbf{x}_{m-1}\\
---\\
 \textbf{x}_{m-2}\\
---\\
 \vdots\\
---\\
 \textbf{x}_1\\
---\\
 \textbf{s}\\
 \end{bmatrix},
\end{equation}
$$~~~~~~(mN_d\times 1)  ~~~~~~~~~~~~~(mN_d\times 1) ~~~~~~~((N_r+N_s)\times 1)$$

\begin{equation}
\textbf{H}_d=\left[
\begin{array}{c c c c c}
\textbf{H}_{r(m-1),d} & & \cdots & & \textbf{0}\\
 & \textbf{H}_{r(m-2),d} & & & \\
\vdots & & \ddots & &\vdots\\
 & & &\textbf{H}_{r(1),d}&\\
\textbf{0} & & \cdots & & \textbf{H}_{s,d}
\end{array}
\right]
\end{equation}
$$~~~(mN_d\times(N_r+N_s))$$

\begin{equation}
\textbf{A}=\left[
\begin{array}{c c c c c}
\textbf{A}_{m-1} & & \cdots & & 0\\
 & \textbf{A}_{m-2} & & & \\
\vdots & & \ddots & &\vdots\\
 & & &\textbf{A}_1&\\
0 & & \cdots & & \textbf{I}
\end{array}
\right]
\end{equation}
$$~~~~~((N_r+N_s)\times(N_r+N_s))$$
Here, we use dashed lines to separate the vectors \textbf{d},
$\textbf{v}_d$ and \textbf{y} in order to distinguish between
transmissions to the destinations in $m$ different time slots.  The
matrix $\textbf{H}_d$ consists of all the channels between each
group of relays and destinations. The matrix \textbf{A} consists
of the amplification coefficients of all relays.

{\color{red}In our transmisstion scheme, all the data packets transmitted from the source nodes and relay nodes contain two parts: a preamble part with training sequence symbols and another part with data symbols. Please see Fig. 3. The source nodes transmit packets and the relay nodes retransmit those packets that contain the identical training sequence symbols which are known at the destination nodes. Therefore, we can make use of them for channel estimation at the destination nodes. After the training sequence, the channel estimation algorithm is switched to decision directed mode \cite{Proakis} and the detected data symbols are fed to the channel estimator. It can continue to estimate and track the channel. Therefore, the channel variation can be tracked after the training phase which can yield better results. Furthermore, this decision directed approach can reduce the length of the training sequence which increases the bandwidth efficiency of the WSNs.}
\begin{figure}[!htb]
\begin{center}
\hspace*{0em}{\includegraphics[width=12cm, height=2cm]{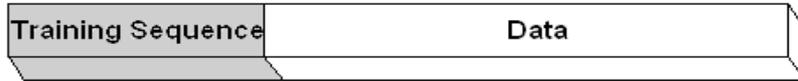}}
\vspace*{-1.0em} \caption{The structure of the packet transmitted
from source nodes and relay nodes}
\end{center}
\end{figure}

\section{Conventional LS and MMSE Channel Estimation}
Consider a channel estimation problem where the output error is defined as:
\begin{equation}
\textbf{e}=\textbf{r}-\textbf{H}\textbf{s}
\end{equation}
where \textbf{s} $(N\times1)$ is the training sequence symbol vector, $\textbf{H}$ $(M\times N)$ is the
estimated channel matrix and \textbf{r} $(M\times1)$ is the received signal
vector at the destination. Conventional channel estimation schemes
seek to find the channel matrix $\textbf{H}$ by minimizing a cost
function which is a suitable objective function of the output error vector \textbf{e}.
\subsection{The LS Channel Estimator}
The least squares (LS) channel estimation minimizes the weighted sum of the squared norm of the error vector $\|\textbf{e}\|^2$ which can be described as:
\begin{equation}
\textbf{H}_{LS}(n)=\arg\min_{\textbf{H}(n)}\sum_{l=1}^{n}\lambda^{n-l}\|\textbf{r}(l)-\textbf{H}(n)\textbf{s}(l)\|^2
\end{equation}
where $\lambda$ denotes the forgetting factor. Computing the gradient of the argument and equating it to a zero matrix, we obtain the LS channel estimator as given by \cite{Kay}:
\begin{equation}
\textbf{H}_{LS}(n)=\left[\sum_{l=1}^{n} \lambda^{n-l}\textbf{r}(l)\textbf{s}^H(l)\right]\left[\sum_{l=1}^{n} \lambda^{n-l}\textbf{s}(l)\textbf{s}^H(l)\right]^{-1}
\end{equation}
where $(\cdot)^H$ and $(\cdot)^{-1}$ denote the complex-conjugate (Hermitian) transpose and the inverse respectively. The LS estimator has a cubic cost with the number of parameters. A complexity reduction is possible by using a recursive procedure that yields the RLS algorithm with quadratic cost.
\subsection{The MMSE Channel Estimator}
The minimum mean square error (MMSE) channel estimation minimizes the expected value of the squared norm of the error vector $\|\textbf{e}\|^2$ which can be described as:
\begin{equation}
\textbf{H}_{MMSE}=\arg\min_{\textbf{H}}E[\|\textbf{r}-\textbf{H}\textbf{s}\|^2]
\end{equation}
After some derivation, the MMSE channel estimator is given by \cite{Kay}:
\begin{equation}
\textbf{H}_{MMSE}=\textbf{R}\left(\textbf{S}^HE[\textbf{H}^H\textbf{H}]\textbf{S}+M\sigma_n^2\textbf{I}\right)^{-1}\textbf{S}^HE[\textbf{H}^H\textbf{H}]
\end{equation}
where $\textbf{S}$ and $\textbf{R}$ are the training sequence symbol matrix and received symbol matrix respectively during a training period. The MMSE channel estimator requires the full a priori knowledge of the channel correlation matrix and the noise variance $\sigma_n^2$ and a cubic cost with the number of parameters.

\section{Set-Membership Channel Estimation}

In contrast with the two conventional channel estimation methods introduced in section III, set-membership (SM)
channel estimation specifies an upper bound $\gamma$ on the norm
of the estimation error vector over a model space of interest
which is denoted as $S$, comprising all possible received
signal pairs $(\textbf{s},\textbf{r})$. The SM criterion
corresponds to finding \textbf{H} that satisfies:
\begin{equation}
\|\textbf{e}(\textbf{H})\|^2\leq\gamma^2,\forall(\textbf{s}, \textbf{r})\in S
\end{equation}
The set of all possible \textbf{H} that satisfy (17) is referred to as the
feasibility set and can be expressed as:
\begin{equation}
\Theta=\bigcap_{(\textbf{s},\textbf{r})\in S}\left\{\textbf{H}\in C^{M\times N}:\|\textbf{r}-\textbf{Hs}\|\leq\gamma\right\}
\end{equation}
At time instant $n$, the constraint set $C_n$  is defined as the
set of all $\textbf{H}(n)$ that satisfy (17) for the
received signal pairs $(\textbf{s}(n),\textbf{r}(n))$:
\begin{equation}
C_n=\left\{\textbf{H}(n)\in C^{M\times N}:\|\textbf{r}(n)-\textbf{H}(n)\textbf{s}(n)\|\leq\gamma\right\}
\end{equation}
The idea behind the SM channel estimation is that if the estimated
channel at a time instant lies outside the
constraint set $C_n$, the estimated channel at the next time instant will lie on the closest boundary of $C_n$. Otherwise,
there is no need to compute and the power consumption can be significantly reduced. This SM approach makes the estimator adapt only in the direction that is necessary.

\subsection{Proposed SM-NLMS Channel Estimation}

The basic update in the LMS Channel Estimation can be written as:
\begin{equation}
\textbf{H}(n+1)=\textbf{H}(n)+\mu(n)\textbf{e}(n)\textbf{s}^H(n)
\end{equation}
where $\textbf{e}(n) = \textbf{r}(n)- \textbf{H}(n)\textbf{s}(n)$ denotes the a priori error vector at time instant $n$, and $\mu(n)$ is the time-dependent step size.  Then we can get a posterior error vector:
\begin{equation}
\textbf{g}(n)=\textbf{r}(n)-\textbf{H}(n+1)\textbf{s}(n)
\end{equation}
By substituting (20) into (21), we have:
\begin{equation}
\begin{split}
\textbf{g}(n)&=\textbf{r}(n)-\left(\textbf{H}(n)+\mu(n)\textbf{e}(n)\textbf{s}^H(n)\right)\textbf{s}(n)\\
&=\left(\textbf{r}(n)-\textbf{H}(n)\textbf{s}(n)\right)-\mu(n)\textbf{e}(n)\textbf{s}^H(n)\textbf{s}(n)\\
&=\textbf{e}(n)-\mu(n)\textbf{e}(n)\textbf{s}^\textbf{H}(n)\textbf{s}(n)
\end{split}
\end{equation}
The constraint set is described as:
\begin{equation}
\|\textbf{g}(n)\|=\|\textbf{e}(n)-\mu(n)\textbf{e}(n)\textbf{s}^H(n)\textbf{s}(n)\|\leq\gamma
\end{equation}
If $\|\textbf{e}(n)\|>\gamma$, then the previous solution lies outside the constraint set. We can choose the constraint value $\|\textbf{g}(n)\|$ equal to $\gamma$ so that the new solution lies on the closest boundary of the constraint set. Therefore:
\begin{equation}
\|\textbf{g}(n)\|=\|\textbf{e}(n)\|\left|1-\mu(n)\textbf{s}^H(n)\textbf{s}(n)\right|=\gamma
\end{equation}
Hence the step size at the $n$th iteration $\mu(n)$ can be expressed as:
\begin{equation}
\mu(n)=\frac{1}{\textbf{s}^H(n)\textbf{s}(n)}\left(1-\frac{\gamma}{\|\textbf{e}(n)\|}\right)
\end{equation}
Finally, we can write the update equation as:
\begin{equation}
\textbf{H}(n+1)=\textbf{H}(n)+\mu(n)\textbf{e}(n)\textbf{s}^H(n)
\end{equation}
where,
\begin{equation}
\mu(n) = \left\{ \begin{array}{ll}
\frac{1}{\textbf{s}^H(n)\textbf{s}(n)}\left(1-\frac{\gamma}{\|\textbf{e}(n)\|}\right), & \textrm{if $\|\textbf{e}(n)\|>\gamma$,}\\0, & \textrm{otherwise.}\\
\end{array}\right.
\end{equation}
Equation (27) shows that the estimated channel matrix updates with a specified step size, only when the norm of the estimation error vector is larger than a fixed error bound which we set. Otherwise, the step sizes are zeros which means there is no update at these time instants.

\subsection{Proposed BEACON Channel Estimation}

The proposed BEACON channel estimation method can be considered as
the following optimization problem:
\begin{equation}
\begin{split}
{\textrm{minimize}} & ~ \sum_{l=1}^{n-1}\lambda(n)^{n-l}\|\textbf{r}(l)-\textbf{H}(n)\textbf{s}(l)\|^2  \\
{\textrm {subject to}} & ~ \|\textbf{r}(n)-\textbf{H}(n)\textbf{s}(n)\|^2 =\gamma^2
\end{split}
\end{equation}
To solve this constrained optimization problem, we can modify the
LS cost function using the method of Lagrange multipliers which
yields the following Lagrangian function:
\begin{equation}
\mathcal{L}=\sum_{l=1}^{n-1}\lambda(n)^{n-l}\|\textbf{r}(l)-\textbf{H}(n)\textbf{s}(l)\|^2+\lambda(n)\left[\|\textbf{r}(n)-\textbf{H}(n)\textbf{s}(n)\|^2-\gamma^2\right]
\end{equation}
where $\lambda(n)$ plays the role of both the Lagrange multiplier
and the forgetting factor of the LS cost function. By setting the
gradient of $\mathcal{L}$ with respect to $\textbf{H}(n)$ equal to zero, after some mathematical manipulations (see Appendix), we get the desired recursive equation for updating the channel matrix $\textbf{H}(n)$:
\begin{equation}
\textbf{H}(n)=\textbf{H}(n-1)+\lambda(n)\boldsymbol{\epsilon}(n)\textbf{k}(n)
\end{equation}
and the recursive equation for updating the intermediate variable matrix $\textbf{P}(n)$:
\begin{equation}
\textbf{P}(n)=\textbf{P}(n-1)-\lambda(n)\textbf{P}(n-1)\textbf{s}(n)\textbf{k}(n)
\end{equation}
where $\boldsymbol{\epsilon}(n)=\textbf{r}(n)-\textbf{H}(n-1)\textbf{s}(n)$
denotes the prediction error vector at time instant $n$. The relationship between $\textbf{k}(n)$ and $\textbf{P}(n-1)$ is
\begin{equation}
\textbf{k}(n)=\frac{\textbf{s}^H(n)\textbf{P}(n-1)}{1+\lambda(n)\textbf{s}^H(n)\textbf{P}(n-1)\textbf{s}(n)}
\end{equation}
The error vector is:
\begin{equation}
\textbf{e}(n)=\textbf{r}(n)-\textbf{H}(n)\textbf{s}(n)
\end{equation}
By substituting (30) into (33), we have:
\begin{equation}
\begin{split}
\textbf{e}(n)&=\textbf{r}(n)-\left[\textbf{H}(n-1)+\lambda(n)\boldsymbol{\epsilon}(n)\textbf{k}(n)\right]\textbf{s}(n)\\
&=\textbf{r}(n)-\textbf{H}(n-1)\textbf{s}(n)-\lambda(n)\boldsymbol{\epsilon}(n)\textbf{k}(n)\textbf{s}(n)\\
&=\boldsymbol{\epsilon}(n)-\lambda(n)\boldsymbol{\epsilon}(n)\frac{\textbf{s}^H(n)\textbf{P}(n-1)\textbf{s}(n)}{1+\lambda(n)\textbf{s}^H(n)\textbf{P}(n-1)\textbf{s}(n)}\\
&=\boldsymbol{\epsilon}(n)-\lambda(n)\boldsymbol{\epsilon}(n)\frac{G(n)}{1+\lambda(n)G(n)}\\
&=\boldsymbol{\epsilon}(n)\left[1-\frac{\lambda(n)G(n)}{1+\lambda(n)G(n)}\right]\\
&=\boldsymbol{\epsilon}(n)\frac{1}{1+\lambda(n)G(n)}
\end{split}
\end{equation}
where $G(n)=\textbf{s}^\textbf{H}(n)\textbf{P}(n-1)\textbf{s}(n)$.
The constraint set is described as:
\begin{equation}
\|\textbf{e}(n)\|=\|\boldsymbol{\epsilon}(n)\frac{1}{1+\lambda(n)G(n)}\|\leq \gamma
\end{equation}
If $\|\boldsymbol{\epsilon}(n)\|>\gamma$, then the previous
solution lies outside the constraint set. We can choose the
constraint value $\|\textbf{e}(n)\|$ equal to $\gamma$ so that the
new solution lies on the closest boundary of the constraint set.
Therefore:
\begin{equation}
\|\textbf{e}(n)\|=\|\boldsymbol{\epsilon}(n)\|\frac{1}{|1+\lambda(n)G(n)|}=\gamma.
\end{equation}
Hence the optimal forgetting factor at the $n$th iteration  can be expressed as:
\begin{equation}
\lambda(n)=\frac{1}{G(n)}\left(\frac{\|\boldsymbol{\epsilon}(n)\|}{\gamma}-1\right)
\end{equation}
Table I shows a summary of the BEACON channel estimation algorithm
which will be used for the simulations.
\begin{table}
  \centering
  \caption{Summary of the BEACON Channel Estimation Algorithm}\label{}
  \begin{tabular}{c}
  \hline
Initialize the algorithm by setting \\
$ {\textbf{H}}(0)=0 $ \\
$ {\textbf{P}}(0)={\textbf{I}}  $ \\
For each instant of time, $n$=1, 2, ..., compute\\
$\boldsymbol{\epsilon}(n)=\textbf{r}(n)-\textbf{H}(n-1)\textbf{s}(n)$\\
$\lambda(n) = \left\{ \begin{array}{ll}
\frac{1}{G(n)}\left(\frac{\|\boldsymbol{\epsilon}(n)\|}{\gamma}-1\right), & \textrm{if $\|\boldsymbol{\epsilon}(n)\|>\gamma$,}\\0, & \textrm{otherwise.}\\
\end{array}\right.
$\\
where $G(n)=\textbf{s}^H(n)\textbf{P}(n-1)\textbf{s}(n)$\\
$\textbf{k}(n)=\frac{\textbf{s}^H(n)\textbf{P}(n-1)}{1+\lambda(n)G(n)}$\\
$\textbf{H}(n)=\textbf{H}(n-1)+\lambda(n)\boldsymbol{\epsilon}(n)\textbf{k}(n)$\\
$\textbf{P}(n)=\textbf{P}(n-1)-\lambda(n)\textbf{P}(n-1)\textbf{s}(n)\textbf{k}(n)$\\
\hline
\end{tabular}
\end{table}

\subsection{Time-Varying Bound}
In order to obtain the optimal error bound at each time instant,
in this section we introduce an error bound function which can
adjust the error bound automatically with the update of the
channel estimate. A similar bound for the SM filtering techniques has been described in \cite{Lamare}. For channel estimation, the bound is heuristic and employs the CSI parameter matrix and the noise variance that should be related with the estimates of interest. It can be expressed as:
\begin{equation}
\gamma(n+1)=(1-\beta)\gamma(n)+\beta\sqrt{\alpha\|\textbf{H}(n)\|^2\sigma^2},
\end{equation}
where $\beta$ is the forgetting factor, $\alpha$ is the tuning
parameter, and $\sigma^2$ is the variance of the noise which is
assumed to be known at the destinations. This time-varying bound is recursive so that it can be used to avoid too high or low values of $\|\textbf{H}(n)\|^2$.

\section{Analysis of the Proposed Algorithms}

\subsection{Steady-State Output MSE Analysis}
In this subsection, we investigate the output MSE in the SM-NLMS and the BEACON channel estimation. The received signal at time instant $n$ is given by:
\begin{equation}
\textbf{r}(n)=\textbf{H}_0\textbf{s}(n)+\textbf{n}(n)
\end{equation}
where $\textbf{H}_0$ $(M\times N)$ is the channel matrix needed to be estimated and $\textbf{n}(n)$ is measurement noise which is assumed here to be Gaussian with zero mean and variance $\sigma_n^2$. Defining the channel estimation error matrix as:
\begin{equation}
\Delta \textbf{H}(n)=\textbf{H}_0-\textbf{H}(n)
\end{equation}
we can express the output error vector as:
\begin{equation}
\begin{split}
\textbf{e}(n)&=\textbf{r}(n)-\textbf{H}(n)\textbf{s}(n)\\
    &=\textbf{r}(n)-[\textbf{H}_0-\Delta \textbf{H}(n)]\textbf{s}(n)\\
    &=\textbf{r}(n)-\textbf{H}_0s(n)+\Delta \textbf{H}(n)\textbf{s}(n)\\
    &=\textbf{n}(n)+\Delta \textbf{H}(n)\textbf{s}(n)
\end{split}
\end{equation}
Therefore, the output MSE expression can be derived as:
\begin{equation}
\begin{split}
J(n)&=E[\|\textbf{e}(n)\|^2]\\
    &=E[\textbf{e}^H(n)\textbf{e}(n)]\\
    &=E\{[\textbf{n}^H(n)+\textbf{s}^H(n)\Delta \textbf{H}^H(n)][\textbf{n}(n)+\Delta \textbf{H}(n)\textbf{s}(n)]\}\\
    &=E[\|\textbf{n}(n)\|^2]+E[\textbf{s}^H(n)\Delta \textbf{H}^H(n)\Delta \textbf{H}(n)\textbf{s}(n)]\\
    &=M\sigma_n^2+E\{tr[\textbf{s}^H(n)\Delta \textbf{H}^H(n)\Delta \textbf{H}(n)\textbf{s}(n)]\}\\
    &=M\sigma_n^2+tr\{E[\textbf{s}^H(n)\Delta \textbf{H}^H(n)\Delta \textbf{H}(n)\textbf{s}(n)]\}
\end{split}
\end{equation}
where $tr(\cdot)$ denotes the trace of a matrix. The property of the matrix trace $tr(\textbf{XY})=tr(\textbf{YX})$ will be used in the following derivation. From (42), we can define the output excess MSE as:
\begin{equation}
\begin{split}
J_{ex}(n)&=tr\{E[\textbf{s}^H(n)\Delta \textbf{H}^H(n)\Delta \textbf{H}(n)\textbf{s}(n)]\}\\
         &=tr\{E[\textbf{s}(n)\textbf{s}^H(n)\Delta \textbf{H}^H(n)\Delta \textbf{H}(n)]\}
\end{split}
\end{equation}

\subsubsection{For the SM-NLMS}
The update equations for the SM-NLMS channel estimation are given by (26) and (27). In (27) $\textbf{s}^H(n)\textbf{s}(n)$ is equal to $N\sigma_s^2$,
where $\sigma_s^2$ is the variance of the pilot signal. By substituting (27) into (26), we can achieve an alternative update equation:
\begin{equation}
\textbf{H}(n+1)=\textbf{H}(n)+\frac{1}{N\sigma_s^2}\left(1-\frac{\gamma}{\|\textbf{e}_0(n)\|}\right)\textbf{e}(n)\textbf{s}^H(n)
\end{equation}
where
\begin{equation}
\|\textbf{e}_0(n)\| = \left\{ \begin{array}{ll}
\|\textbf{e}(n)\|, & \textrm{if $\|\textbf{e}(n)\|>\gamma$,}\\ \gamma, & \textrm{otherwise.}\\
\end{array}\right.
\end{equation}
As a consequence, the update equation of the channel estimation error can be expressed as:
\begin{equation}
\begin{split}
\Delta \textbf{H}(n+1)=&\Delta \textbf{H}(n)-\frac{1}{N\sigma_s^2}\left(1-\frac{\gamma}{\|\textbf{e}_0(n)\|}\right)\textbf{e}(n)\textbf{s}^H(n)\\
             =&\Delta \textbf{H}(n)-\frac{1}{N\sigma_s^2}\textbf{e}(n)\textbf{s}^H(n)
              +\frac{\gamma}{N\sigma_s^2}\frac{\textbf{e}(n)}{\|\textbf{e}_0(n)\|}\textbf{s}^H(n)
\end{split}
\end{equation}
Then, we can use (46) to derive the update equation of the output excess MSE in (43) (see Appendix):
\begin{equation}
J_{ex}(n+1)=M\sigma_n^2+2\gamma E\left[\frac{1}{\|\textbf{e}_0(n)\|}\right]J_{ex}(n)
            -2\gamma E\left[\frac{\|\textbf{e}(n)\|^2}{\|\textbf{e}_0(n)\|}\right]+\gamma^2 E\left[\frac{\|\textbf{e}(n)\|^2}{\|\textbf{e}_0(n)\|^2}\right]
\end{equation}
From (45), the three expected values in (47) can be expressed as:
\begin{equation}
E\left[\frac{1}{\|\textbf{e}_0(n)\|}\right]=E\left[\frac{1}{\|\textbf{e}(n)\|}\bigg|\|\textbf{e}(n)\|>\gamma \right]P_{up}+\frac{1}{\gamma}(1-P_{up})
\end{equation}
\begin{equation}
E\left[\frac{\|\textbf{e}(n)\|^2}{\|\textbf{e}_0(n)\|}\right]=E\left[\|\textbf{e}(n)\|\big|\|\textbf{e}(n)\|>\gamma\right]P_{up}
                                        +\frac{1}{\gamma}E\left[\|\textbf{e}(n)\|^2\big|\|\textbf{e}(n)\|\leq\gamma\right](1-P_{up})
\end{equation}
\begin{equation}
E\left[\frac{\|\textbf{e}(n)\|^2}{\|\textbf{e}_0(n)\|^2}\right]=P_{up}+\frac{1}{\gamma^2}E\left[\|\textbf{e}(n)\|^2\big|\|\textbf{e}(n)\|\leq\gamma\right](1-P_{up})
\end{equation}
where $E[\cdot\big|\cdot]$ denotes the conditional expected value and $P_{up}$ stands for the probability of update in each recursion.
Let:
\begin{equation}
X_1=E\left[\frac{1}{\|\textbf{e}(n)\|}\bigg|\|\textbf{e}(n)\|>\gamma \right]
\end{equation}
\begin{equation}
Y_1=E\left[\|\textbf{e}(n)\|\big|\|\textbf{e}(n)\|>\gamma\right]
\end{equation}
\begin{equation}
Z_1=E\left[\|\textbf{e}(n)\|^2\big|\|\textbf{e}(n)\|\leq\gamma\right]
\end{equation}
Equation (47) becomes:
\begin{equation}
\begin{split}
J_{ex}(n+1)=&M\sigma_n^2+[2\gamma X_1P_{up}+2(1-P_{up})]J_{ex}(n)-2\gamma Y_1P_{up}
            -2Z_1(1-P_{up})+\gamma^2P_{up}+Z_1(1-P_{up})\\
           =&[2\gamma X_1P_{up}+2-2P_{up}]J_{ex}(n)-2\gamma Y_1P_{up}
            -Z_1(1-P_{up})+M\sigma_n^2+\gamma^2P_{up}
\end{split}
\end{equation}
During the steady state, $J_{ex}(n+1)\rightarrow J_{ex}(n)$. Therefore, the steady-state output excess MSE expression of the SM-NLMS channel estimation is:
\begin{equation}
J_{ex}(n)=\frac{2\gamma Y_1P_{up}+Z_1(1-P_{up})-M\sigma_n^2-\gamma^2P_{up}}{2\gamma X_1P_{up}-2P_{up}+1}
\end{equation}

\subsubsection{For the BEACON}
According to Table I, we can get the update equation of the channel estimation error for the BEACON channel estimation which is very similar to (46):
\begin{equation}
\Delta \textbf{H}(n)=\Delta \textbf{H}(n-1)-\frac{\boldsymbol{\epsilon}(n)\textbf{s}^H(n)\textbf{P}(n-1)}{G(n)}
            +\gamma\frac{\boldsymbol{\epsilon}(n)}{\|\boldsymbol{\epsilon}_0(n)\|}\frac{\textbf{s}^H(n)\textbf{P}(n-1)}{G(n)}
\end{equation}
where,
\begin{equation}
\|\boldsymbol{\epsilon}_0(n)\| = \left\{ \begin{array}{ll}
\|\boldsymbol{\epsilon}(n)\|, & \textrm{if $\|\boldsymbol{\epsilon}(n)\|>\gamma$,}\\ \gamma, & \textrm{otherwise.}\\
\end{array}\right.
\end{equation}
Following the same steps described for the SM-NLMS channel estimation in the Appendix, we find that the steady-state output excess MSE expression of the BEACON channel estimation has the same style as (55):
\begin{equation}
J_{ex}(n)=\frac{2\gamma Y_2P_{up}+Z_2(1-P_{up})-M\sigma_n^2-\gamma^2P_{up}}{2\gamma X_2P_{up}-2P_{up}+1}
\end{equation}
where,
\begin{equation}
X_2=E\left[\frac{1}{\|\boldsymbol{\epsilon}(n)\|}\bigg|\|\boldsymbol{\epsilon}(n)\|>\gamma \right]
\end{equation}
\begin{equation}
Y_2=E\left[\|\boldsymbol{\epsilon}(n)\|\big|\|\boldsymbol{\epsilon}(n)\|>\gamma\right]
\end{equation}
\begin{equation}
Z_2=E\left[\|\boldsymbol{\epsilon}(n)\|^2\big|\|\boldsymbol{\epsilon}(n)\|\leq\gamma\right]
\end{equation}

\subsubsection{The Probability of Update $P_{up}$}
From (27), we can get the relation about the probability of update of the SM-NLMS channel estimation:
\begin{equation}
P_{up}=Pr\{\|\textbf{e}(n)\|>\gamma\}=Pr\{\|\textbf{e}(n)\|^2>\gamma^2\}
\end{equation}
Similarly, for the BEACON channel estimation we just need to use $\boldsymbol{\epsilon}(n)$ instead of $\textbf{e}(n)$. It is easy to see that $P_{up}$ depends on the distribution of $\|\textbf{e}(n)\|^2$.
For the estimated channel matrix $\textbf{H}_0$ with size $M\times N$:
\begin{equation}
\begin{split}
\|\textbf{e}(n)\|^2&=\sum_{i=1}^{M}(\mathfrak{R}^2[e_i(n)]+\mathfrak{I}^2[e_i(n)])\\
          &=\frac{\sigma_n^2}{2}\sum_{i=1}^{M}(\frac{\mathfrak{R}^2[e_i(n)]}{\sigma_n^2/2}+\frac{\mathfrak{I}^2[e_i(n)]}{\sigma_n^2/2})
\end{split}
\end{equation}
During the steady state, assuming $\Delta \textbf{H}(n)\rightarrow0$, the linear relationship between $\textbf{e}(n)$,$\Delta \textbf{H}(n)$ and $\textbf{n}(n)$ in (41) shows that the distribution of $\textbf{e}(n)$ is typically Gaussian unless a jamming signal with another distribution is present. Therefore we can get that the elements of the error vector $\textbf{e}(n)$ have the same distribution with the elements of the noise vector $\textbf{n}(n)$. Recalling that $\mathfrak{R}[n_i(n)]$ and $\mathfrak{I}[n_i(n)]$ $\sim\mathcal{N}(0,\frac{\sigma_n^2}{2})$, we can express the distribution of (63) by a chi-square random variable with $2M$ degree of freedom as follows:
\begin{equation}
\|\textbf{e}(n)\|^2\sim \frac{\sigma_n^2}{2}\mathcal{X}_{2M}^2
\end{equation}
Therefore, (62) becomes:
\begin{equation}
\begin{split}
P_{up}=&Pr\left\{\sum_{i=1}^{M}(\frac{\mathfrak{R}^2[e_i(n)]}{\sigma_n^2/2}+\frac{\mathfrak{I}^2[e_i(n)]}{\sigma_n^2/2})>\gamma^2\frac{2}{\sigma_n^2}\right\}\\
      =&1-Pr\left\{\sum_{i=1}^{M}(\frac{\mathfrak{R}^2[e_i(n)]}{\sigma_n^2/2}+\frac{\mathfrak{I}^2[e_i(n)]}{\sigma_n^2/2})\leq\gamma^2\frac{2}{\sigma_n^2}\right\}\\
      =&1-F\left(\gamma^2\frac{2}{\sigma_n^2};2M\right)
\end{split}
\end{equation}
where $F(\cdot)$ is the chi-square cumulative distribution function (CDF) \cite{Papoulis} defined by:
\begin{equation}
F(x;l)=\frac{\Gamma_L(l/2,x/2)}{\Gamma(l/2)}
\end{equation}
In (66) $\Gamma_L(s,x)$ is the lower incomplete Gamma function:
\begin{equation}
\Gamma_L(s,x)=\int_0^x t^{s-1}e^{-t}dt
\end{equation}
and $\Gamma(x)$ is the gamma function:
\begin{equation}
\Gamma(x)=\int_0^\infty t^{x-1}e^{-t}dt
\end{equation}
By substituting (67) and (68) into (66), we can finally obtain:
\begin{equation}
F(x;l)=\frac{\int_0^{\frac{x}{2}} t^{\frac{l}{2}-1}e^{-t}dt}{\int_0^\infty t^{\frac{l}{2}-1}e^{-t}dt}
\end{equation}
where $l$ denotes the number of degrees of freedom.
\subsection{Computational Complexity Analysis}

Table II lists the computational complexity per update in terms of the number of multiplications, additions and divisions for the SM-NLMS and BEACON algorithms and their competing algorithms. The size of the estimated channel matrix is $M\times N$. For our cooperative WSN system model, when $\textbf{H}_d$ is chosen as the estimated channel, we can get:
\begin{equation}
M=mN_d
\end{equation}
\begin{table}
\begin{scriptsize}
  \centering
  \caption{Computational Complexity per Update}\label{}
  \begin{tabular}{c c c c}
  \hline
  Algorithm & Multiplication & Addition & Division\\
  \hline
  NLMS & $2MN+N+min\{M,N\}$ & $2MN+N-1$ & 1\\
  SM-NLMS & $MN+M+P_{up}(MN+N+min\{M,N\})$ & $MN+M-1+P_{up}(MN+N)$ & 2\\
  RLS & $4N^2+2MN+N$ & $3N^2+2MN-N$ & 2\\
  BEACON & $N^2+MN+M+N+P_{up}(2N^2+MN+N+min\{M,N\})$ & $N^2+MN+M-2+P_{up}(2N^2+MN-N+2)$ & 2\\
  \hline
\end{tabular}
\end{scriptsize}
\end{table}
and,
\begin{equation}
N=N_r+N_s
\end{equation}
Because the multiplication dominates the computational complexity of the algorithms, in order to compare the computational complexity of our proposed algorithms with their competition algorithms,  the number of multiplications versus the size of the channel matrix performance for each update is displayed in Fig. 4. For the purpose of illustration, we set M equal to N. It can be seen that our proposed SM-NLMS and BEACON channel estimation algorithms  have a significant complexity reduction compared with the conventional NLMS and RLS channel estimation algorithms. Obviously, a lower $P_{up}$ will cause a lower computational complexity. Furthermore, assuming the linear MMSE detectors are used in the destination nodes which require cubic complexity, we can get the conclusion that the power used for our proposed channel estimation is only a small fraction of the power budget of these nodes.
\begin{figure}[!htb]
\begin{center}
\def\epsfsize#1#2{0.75\columnwidth}
\epsfbox{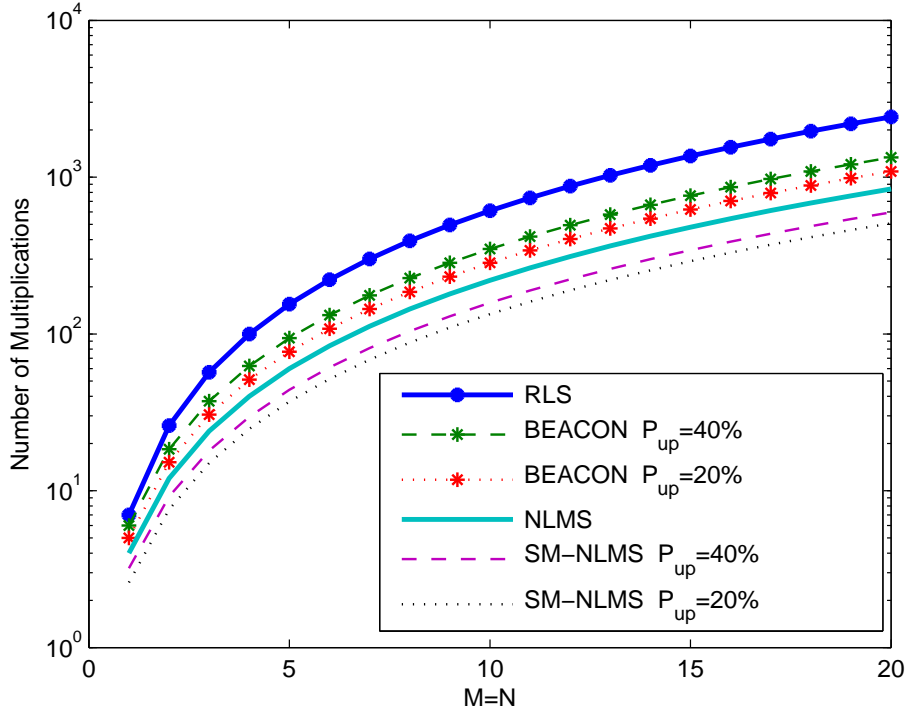} \caption{The number of multiplications versus the size of the channel matrix.}
\end{center}
\end{figure}

\section{Simulations}
In this section, we numerically study the performance of our two proposed SM estimation methods as well as the design of the optimal error bound. We consider a 3-hop ($m$=3) wireless sensor network. The number of sources ($N_s$), two groups of relays ($N_{r(1)}, N_{r(2)}$) and destinations ($N_d$) are 2, 4, 4, 3 respectively. We consider an AF cooperation protocol and the amplification coefficient of each relay is set to 1 for the purpose of simplification. We choose $\textbf{H}_d$ as our estimated channel because it is the most significant and most complex channel among all channels of the WSN system. {\color{red}The quasi-static fading channel (block fading) is considered in our simulations whose elements are Rayleigh random variables and assumed to be invariant during the transmission of each packet. Also, in order to test our proposed channel estimation algorithms in a time-varying environment, we consider a typical fading channel for wireless communications systems, a Rayleigh fading channel, which can be modeled according to Clarke's Model \cite{Rappaport}. According to the transmission scheme introduced in Section II, during each phase, the sources and each group of relays transmit the QPSK modulated packets with $n_p$ symbols among which $n_t$ are training symbols and $n_d$ are data symbols (Note that $n_p=n_t+n_d$). $n_p$, $n_t$ and $n_d$ will be specified in the following simulations.} The noise at the destinations is modeled as circularly symmetric complex Gaussian random variables with zero mean. The SNR is fixed at 10 dB.

\subsection{MSE performance}
Fig. 5 and Fig. 6 show the channel matrix mean square error (MSE) performance of our proposed SM-NLMS and BEACON channel estimation methods for the quasi-static fading channel, and compare them with the conventional NLMS and RLS channel estimation algorithms. For the SM-NLMS estimator, we choose five fixed error bounds ($\gamma$) ranging from 0.3 to 1.1. It can be seen that increasing the error bound makes the update rate (UR) decrease. It means the update is selective which can reduce the computational complexity and power consumption. In the case of an error bound equal to 1.1, the UR can fall dramatically to 0.0868. The optimal error bound appears between 0.7 and 0.9. In that situation, the SM-NLMS channel estimation achieves the fastest convergence speed and lowest steady states. Otherwise, the performance degrades due to overbounding or underbounding. For the BEACON estimator, we choose four fixed error bounds ranging from 0.6 to 0.9. Also, the minimum-mean-square error (MMSE) channel estimator which requires the full a priori knowledge of the channel correlation matrix and the noise variance is used here for reference. It can be seen that a higher value of $\gamma$ results in worse MSE performance but a lower UR. In the case of an error bound equal to 0.6, the BEACON algorithm outperforms the conventional RLS algorithm (with a forgetting factor of 0.998) in terms of convergence speed and steady state with a slightly reduced UR (0.9128). When the error bound is increased to 0.8, although its convergence speed is slower than RLS channel estimation, the final MSE is comparable with a much lower UR (0.4356).
\begin{figure}[!htb]
\begin{center}
\def\epsfsize#1#2{0.75\columnwidth}
\epsfbox{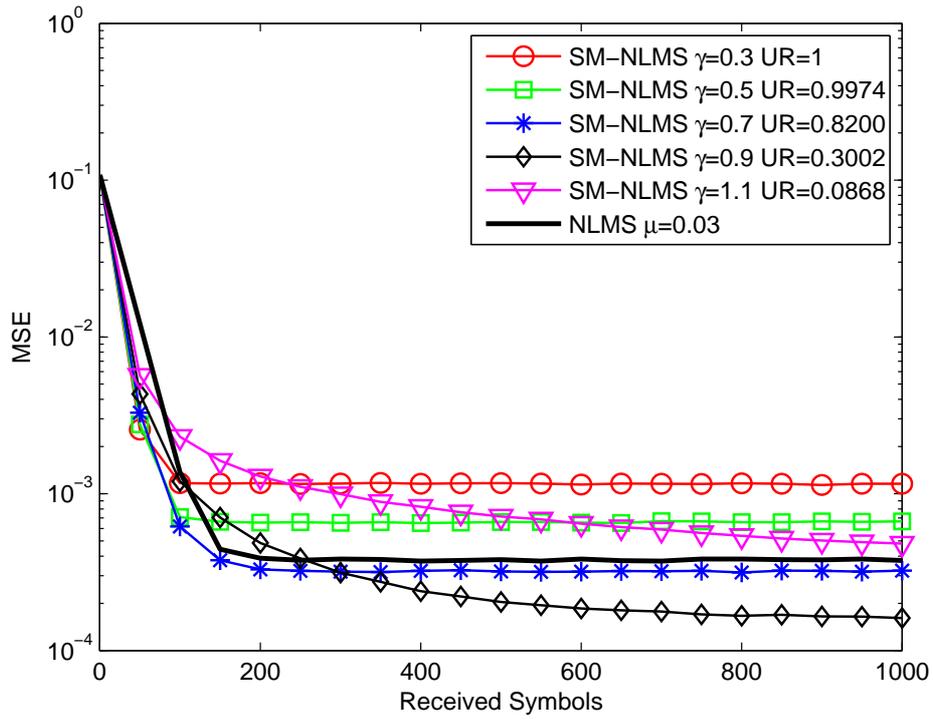} \caption{MSE performance of the SM-NLMS channel
estimation of $\textbf{H}_d$ for quasi-static fading channel compared with the NLMS channel
estimation. $n_p$=1000, $n_t$=100 and $n_d$=900.}
\end{center}
\end{figure}

\begin{figure}[!htb]
\begin{center}
\def\epsfsize#1#2{0.75\columnwidth}
\epsfbox{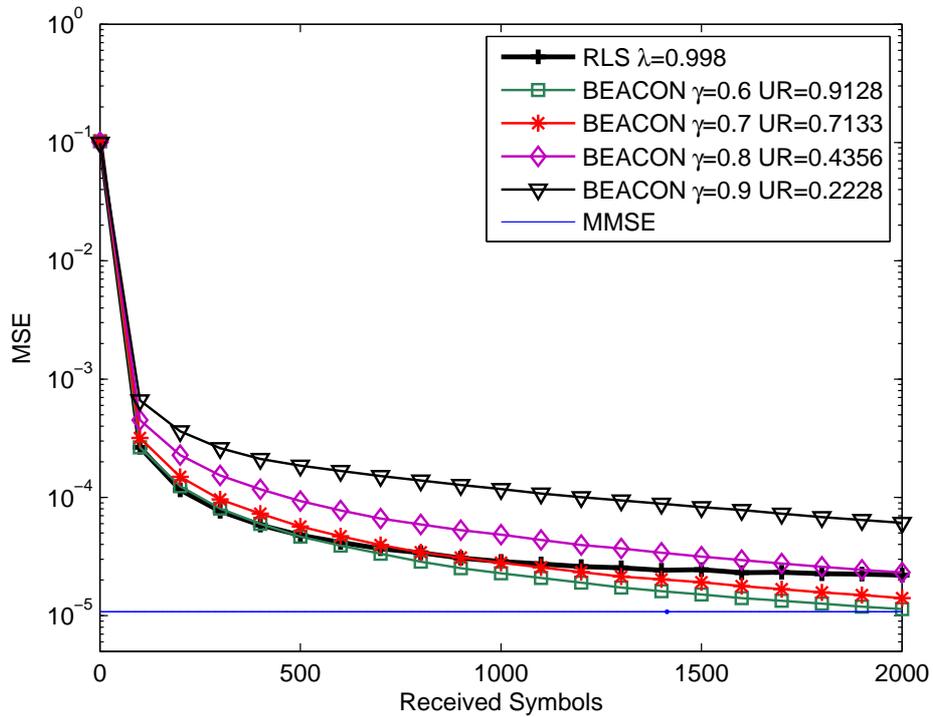} \caption{MSE performance of the BEACON channel
estimation of $\textbf{H}_d$ for quasi-static fading channel compared with the RLS channel
estimation. $n_p$=2000, $n_t$=100 and $n_d$=1900.}
\end{center}
\end{figure}

Fig. 7 and Fig. 8 illustrate the performance when we apply the
time-varying bound (TVB) into the SM-NLMS and BEACON channel
estimation. For the SM-NLMS estimator, we transmit packets with 1000
($n_p$) symbols among which 100 ($n_t$) are training symbols and 900
($n_d$) are data symbols. We set $\alpha$ to 1.5 and $\beta$ to
0.01. The curve of our proposed algorithm lies on the optimal
position which is very close to the curve of the SM-NLMS with fixed
error bound 0.8. Also, its update rate decreases further which is
our expectation. For the BEACON estimator, we transmit packets with
2000 ($n_p$) symbols among which 100 ($n_t$) are training symbols
and 1900 ($n_d$) are data symbols. We set $\alpha$ to 3 and $\beta$
to 0.001. Our proposed algorithm can achieve very similar
performance to the conventional RLS channel estimation with a
substantial reduction in the UR. Therefore, the computational
complexity is significantly reduced.

\begin{figure}[!htb]
\begin{center}
\def\epsfsize#1#2{0.75\columnwidth}
\epsfbox{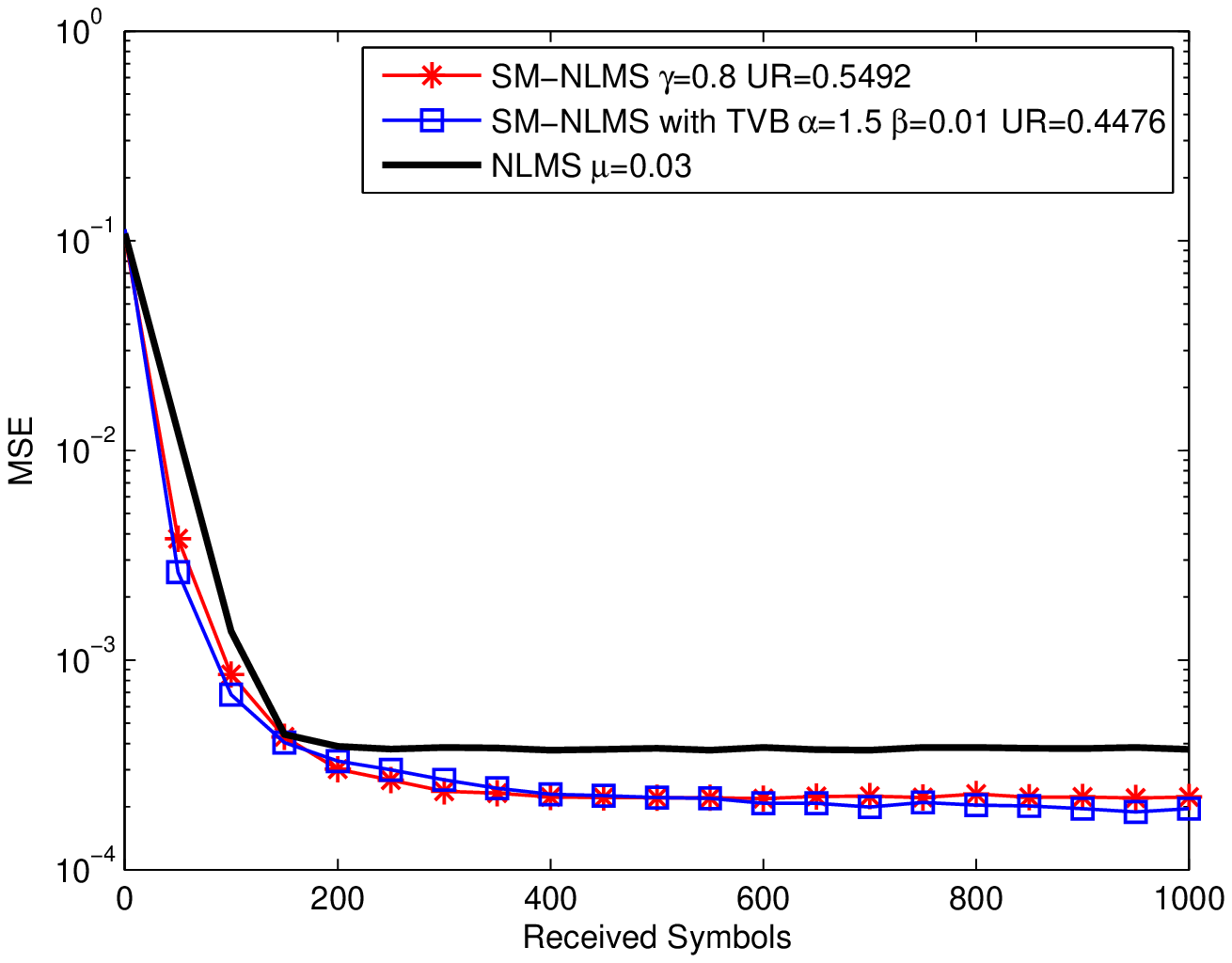} \caption{MSE performance of the SM-NLMS channel
estimation with a time-varying bound for quasi-static fading channel. $n_p$=1000, $n_t$=100 and $n_d$=900.}
\end{center}
\end{figure}

\begin{figure}[!htb]
\begin{center}
\def\epsfsize#1#2{0.75\columnwidth}
\epsfbox{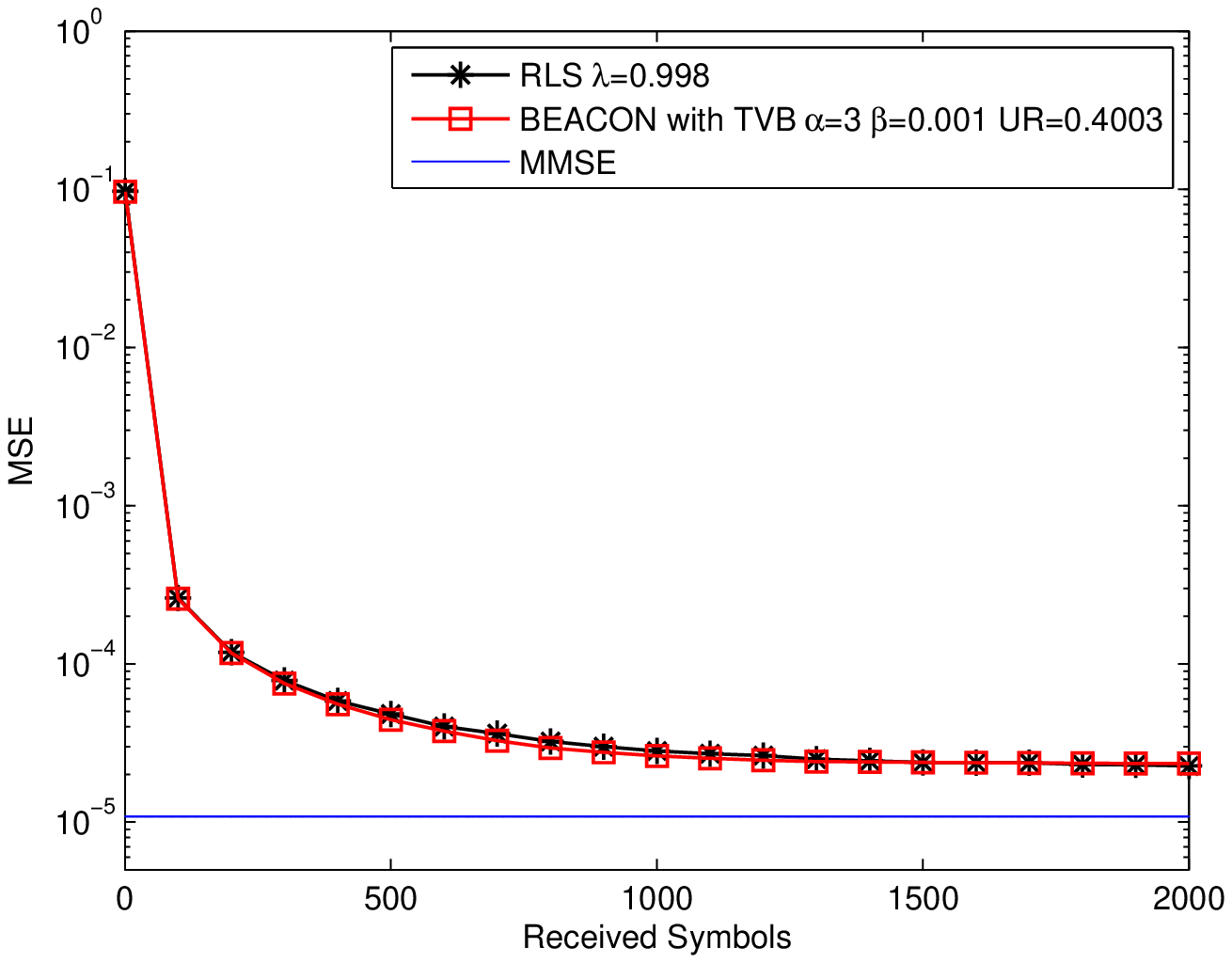} \caption{MSE performance of the BEACON channel
estimation with a time-varying bound for quasi-static fading channel. $n_p$=2000, $n_t$=100 and $n_d$=1900.}
\end{center}
\end{figure}

The MSE versus SNR performance of the SM-NLMS and BEACON channel
estimation methods are displayed with fixed error bounds and the
proposed time-varying error bounds in Fig. 9 and Fig. 10. In the
cases of fixed error bounds, the MSE is lower bounded at different
values for different error bounds. For the SM-NLMS estimator, a
higher SNR needs a specified lower error bound to achieve the
optimal MSE performance.  When the time-varying error bound is
applied, the MSE remains very close to the optimal values for all
SNRs. For the BEACON estimator, when the SNR is larger than a
specified value, its MSE will become worse. However, when the
time-varying error bound (TVB) is applied, it can be observed that
the MSE keeps on decreasing alone with the increase of the SNR.
These two figures show the robustness to the SNR variation of our
proposed algorithms for the quasi-static fading channel.

\begin{figure}[!htb]
\begin{center}
\def\epsfsize#1#2{0.8\columnwidth}
\epsfbox{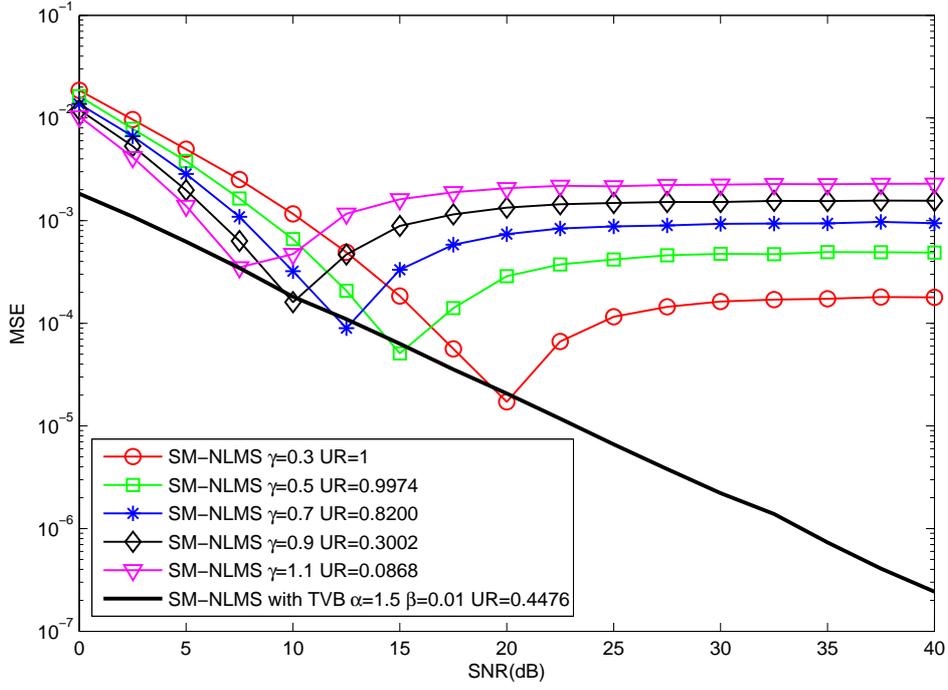} \caption{SM-NLMS channel estimation MSEs versus SNR for both the fixed bound and time-varying bound for quasi-static fading channel. $n_p$=1000, $n_t$=100 and $n_d$=900.}
\end{center}
\end{figure}

\begin{figure}[!htb]
\begin{center}
\def\epsfsize#1#2{0.75\columnwidth}
\epsfbox{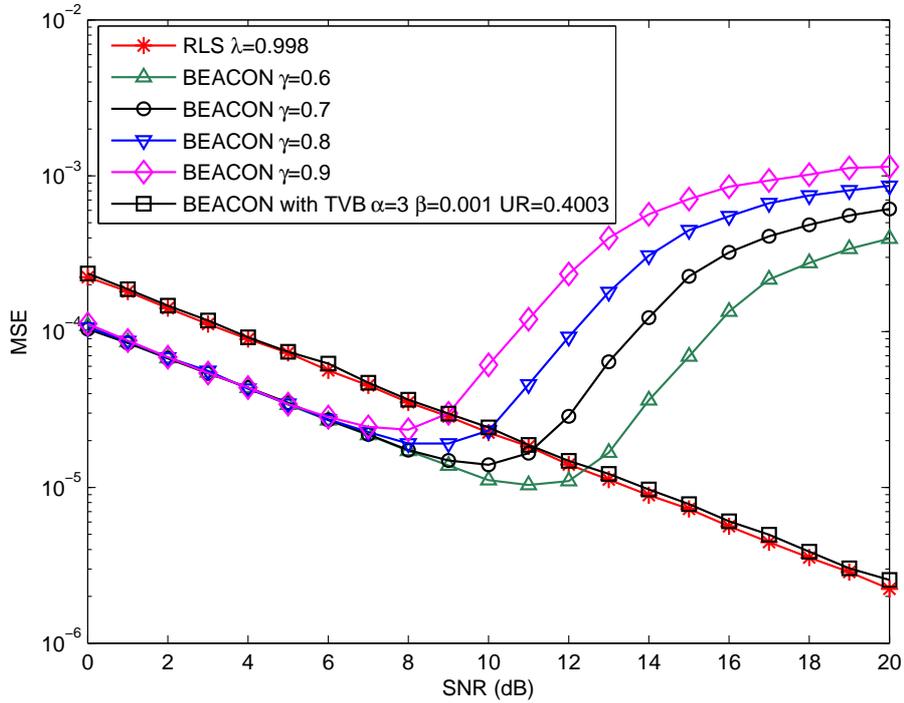} \caption{BEACON channel estimation MSEs versus SNR for both the fixed bound and time-varying bound for quasi-static fading channel. $n_p$=2000, $n_t$=100 and $n_d$=1900.}
\end{center}
\end{figure}
{\color{red}In order to test our proposed channel estimation algorithms in a time-varying environment, we consider a typical fading channel for wireless systems, a Rayleigh fading channel, which can be modeled according to Clarke's Model \cite{Rappaport}. Fig.11 and Fig. 12 show the MSE performance of our proposed channel estimation algorithms for the time-varying fading channel and three different fading rates (normalized Doppler frequency $f_dT$) are used in the simulations: $10^{-5}$, $5\times10^{-5}$, and $10^{-4}$. Because of the requirements of low power consumption and the fact that a fast convergence speed of the proposed algorithms might help reducing the need for long training sequences for the WSNs, we focus on the performance of packets with 500 ($n_p$) symbols among which 50 ($n_t$) are training symbols and 450 ($n_d$) are data symbols. For the SM-NLMS estimator, our proposed algorithm can achieve better performance than the conventional NLMS algorithms for all the three fading rates. Along with the increase of the fading rate, the advantage becomes less pronounced and the update rate becomes higher. For the BEACON estimator, our proposed algorithm can achieve very similar performance to the conventional RLS algorithms for all the three fading rate. (Note that for the conventional RLS algorithms, when increasing the fading rate, we have to lower the forgetting factor to get the optimal performance.) Along with the increasing of the fading rate, the update rate becomes higher. Therefore, we can conclude that our proposed channel estimation algorithms can work well for the time-varying fading channel and for a range of values of $f_dT$.}
\begin{figure}[!htb]
\begin{center}
\def\epsfsize#1#2{0.75\columnwidth}
\epsfbox{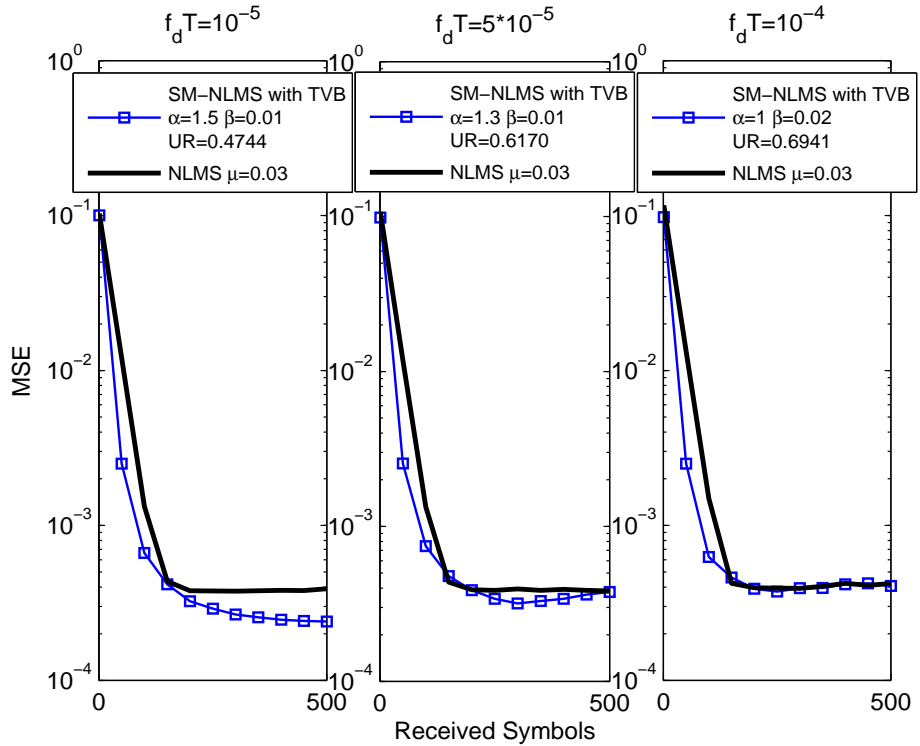} \caption{MSE performance of the SM-NLMS channel
estimation for Rayleigh fading channels compared with the NLMS
channel estimation. $n_p$=500, $n_t$=50 and $n_d$=450.}
\end{center}
\end{figure}

\begin{figure}[!htb]
\begin{center}
\def\epsfsize#1#2{0.75\columnwidth}
\epsfbox{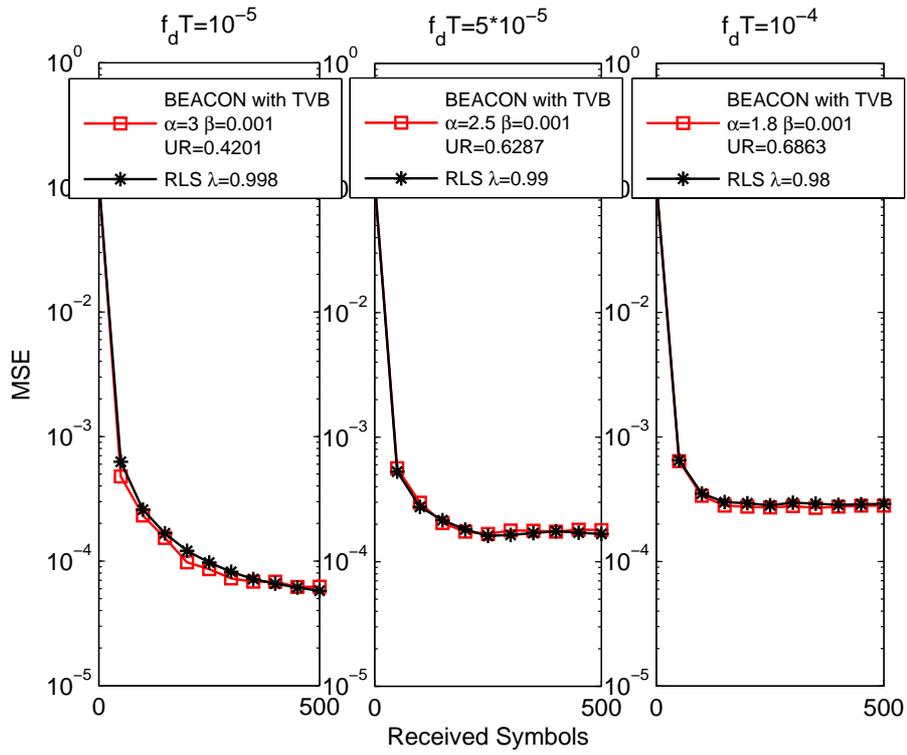} \caption{MSE performance of the BEACON channel
estimation for Rayleigh fading channels compared with the RLS
channel estimation. $n_p$=500, $n_t$=50 and $n_d$=450.}
\end{center}
\end{figure}
\subsection{BER performance}
The MSE performance is very useful to give designers an idea of how well channel estimators perform, whereas bit error rate (BER) performance is meaningful in practice. Therefore, in this subsection we focus on the BER performance of our proposed algorithms. We consider a simulation where the data packets transmitted at the sources nodes have 1000 ($n_p$) symbols and trained with 100 ($n_t$) symbols. Linear MMSE detectors are used in the destination nodes. We choose $\textbf{H}_d$ as our estimated channel and other channels are assumed to be known. Quasi-static fading channel are considered. It can be seen from Fig. 13 that our two proposed SM channel estimation algorithms with time varying bound can achieve a similar BER performance to their competing algorithms. Also, the BEACON channel estimator has lower BER than the SM-NLMS channel estimator due to the higher computational complexity and the use of the second-order statistics.
\begin{figure}[!htb]
\begin{center}
\def\epsfsize#1#2{0.75\columnwidth}
\epsfbox{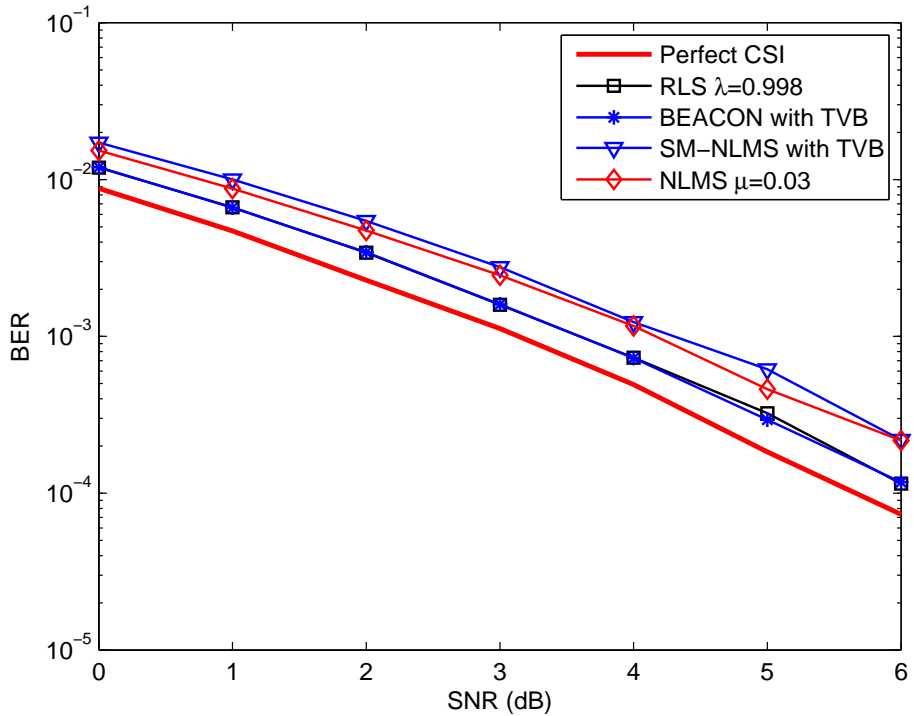} \caption{BER performance of the proposed channel estimation algorithms. $n_p$=1000, $n_t$=100 and $n_d$=900.}
\end{center}
\end{figure}

\subsection{Verification of the Analysis}
In this subsection, experiments were conducted to validate our analysis of the SM-NLMS and BEACON algorithms. From (70) and (71), the two variables $M$ and $N$ using in the section V can be obtained. $M$=9 and $N$=10. First of all, the analysis of the probability of update is verified using (65). It can be seen from Fig. 14 that the $P_{up}$ in simulations of the SM-NLMS and BEACON channel estimation is close to and lower bounded by the $P_{up}$ from our analysis. The gap between the analytical curve and the simulations of two SM channel estimation is due to the approximation made in the analysis. In section V, we assume that the channel matrix error $\Delta \textbf{H}$ approaches zero during the steady-state. However, for the SM algorithms it is not accurate because the bound set for the output estimation error would enlarge the $\Delta \textbf{H}$. During the steady-state, the SM-NLMS channel estimation has a larger $\Delta \textbf{H}$ than the BEACON channel estimation which therefore causes a larger gap between the analytical curve and the simulation. After that we continue to verify the analysis of the steady-state output excess MSE using (55) and (58). Because it is difficult to obtain the full-analytical expressions of the conditional expected values $X_1, Y_1, Z_1, X_2, Y_2, Z_2$, a semi-analytical method is used here. It means that the data from the simulations is used to calculate these conditional expected values in (55) and (58). In order to lower the effect of the difference between the analytical $P_{up}$ and the simulation $P_{up}$ of the SM-NLMS channel estimation, $1.1\sigma_n^2$ is chosen approximately to take the place of $\sigma_n^2$ in (65) which would produce a more accurate $\Delta \textbf{H}$ and $P_{up}$ for the SM-NLMS channel estimation.  Fig. 15 and Fig. 16 show the steady-state output excess MSE versus $\gamma^2/(mN_d\sigma_n^2)$ of the two channel estimation algorithms. From the figures, it can be seen that the semi-analytical curves can match the simulation curves well. Therefore, it can be stated that our analysis is able to predict accurately the output steady-state excess MSE for different choices of bound $\gamma$.
\begin{figure}[!htb]
\begin{center}
\def\epsfsize#1#2{0.75\columnwidth}
\epsfbox{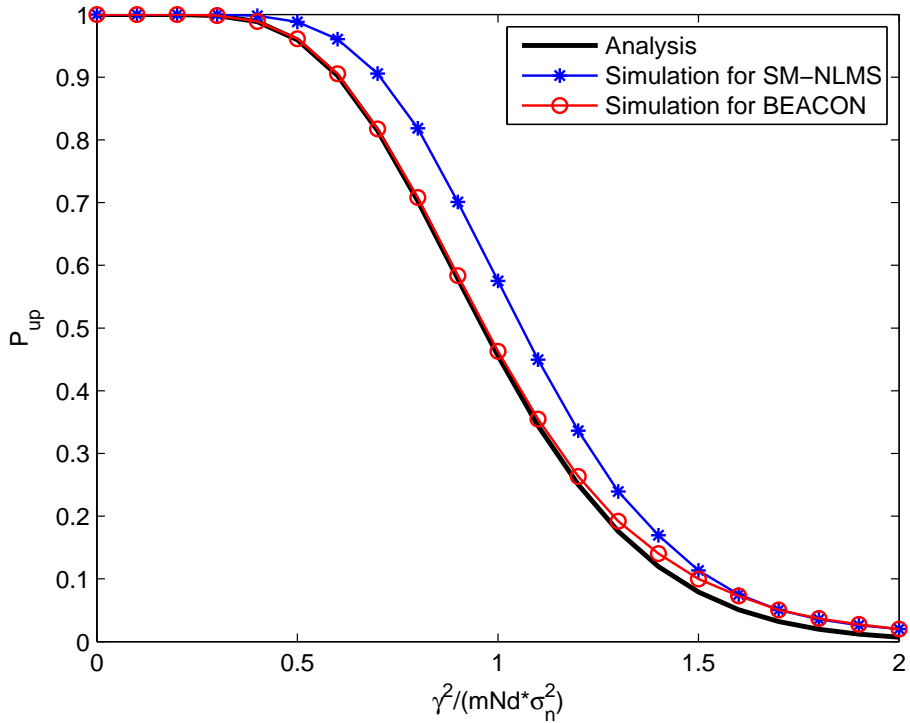} \caption{Analysis of the probability of the
update $P_{up}$.}
\end{center}
\end{figure}

\begin{figure}[!htb]
\begin{center}
\def\epsfsize#1#2{0.75\columnwidth}
\epsfbox{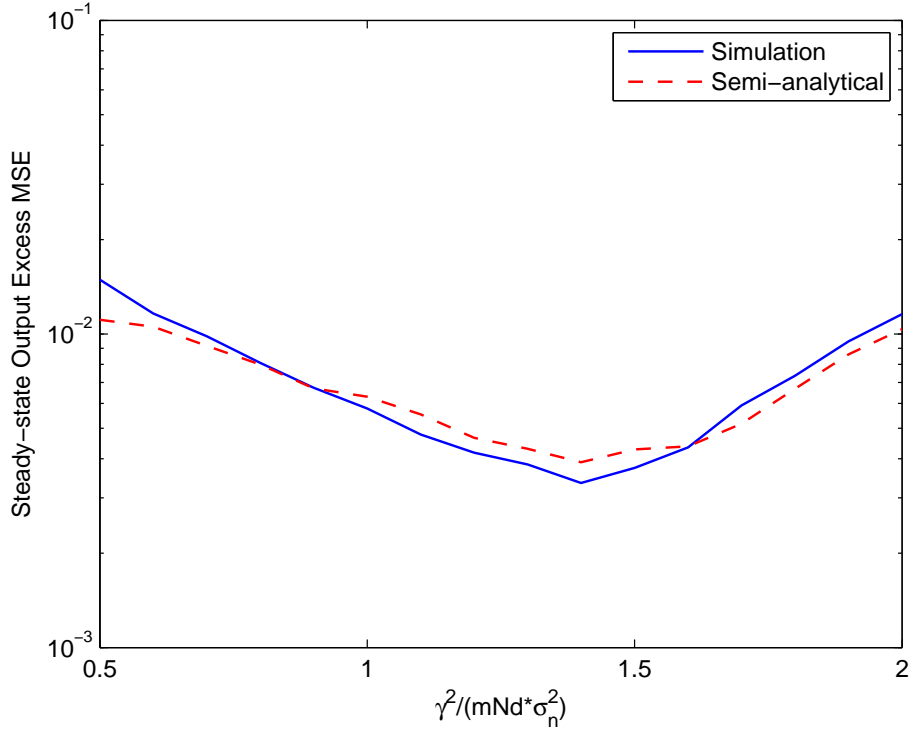} \caption{Steady-state excess MSE analysis for the SM-NLMS channel estimation.}
\end{center}
\end{figure}

\begin{figure}[!htb]
\begin{center}
\def\epsfsize#1#2{0.75\columnwidth}
\epsfbox{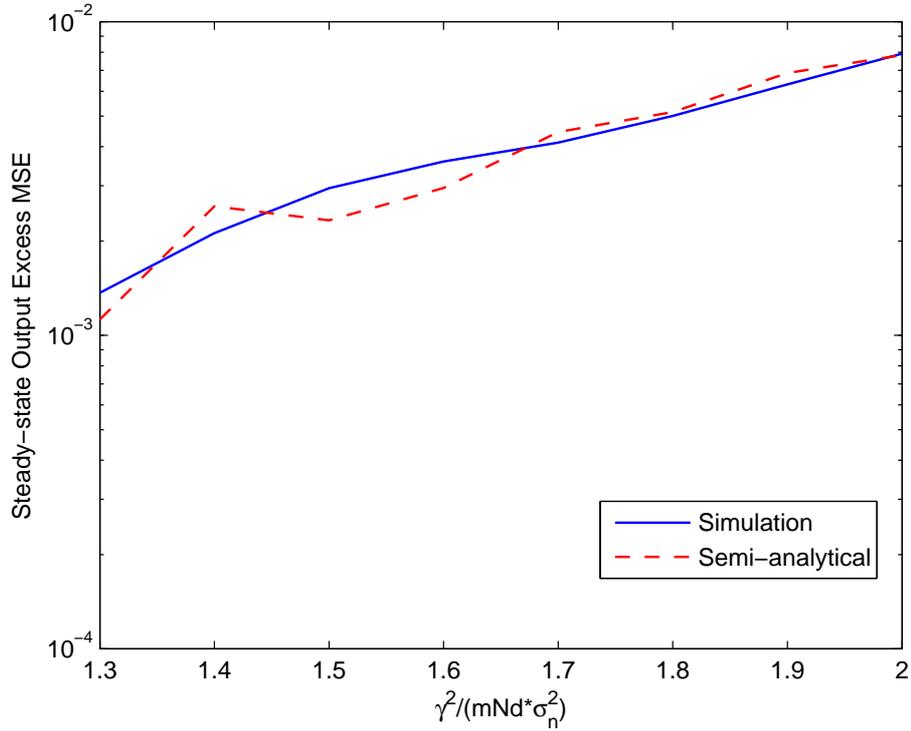} \caption{Steady-state excess MSE analysis for the BEACON channel estimation..}
\end{center}
\end{figure}

\section{Conclusions}
Two SM channel estimation methods have been proposed based on time-varying bound for cooperative wireless sensor networks. It has been shown that our proposed methods can achieve better or similar performance to conventional NLMS and RLS channel estimation, offering reduced computational complexity. Analyses of the steady-state MSE and computational complexity are presented for the two channel estimation and closed-form expressions of the excess MSE and the probability of update are provided. Furthermore, the incorporation of the time-varying bound function makes it robust to changes in the environment. These features are desirable for WSNs and bring about a significant reduction in energy consumption.

\appendix

\section*{part of the derivations about the proposed BEACON channel estimation algorithm}
By setting the gradient of $\mathcal{L}$ in (29) with respect to $\textbf{H}(n)$ equal to zero, we have
\begin{equation}
\frac{\partial \mathcal{L}}{\partial \textbf{H}(n) }=2\sum_{l=1}^{n-1}\lambda(n)^{n-l}\left[\textbf{r}(l)-\textbf{H}(n)\textbf{s}(l)\right][-\textbf{s}^H(l)]+2\lambda(n)[\|\textbf{r}(n)-\textbf{H}(n)\textbf{s}(n)][-\textbf{s}^H(n)]=0
\end{equation}
Therefore,
\begin{equation}
\textbf{H}(n)\left[\sum_{l=1}^{n-1} \lambda(n)^{n-l}\textbf{s}(l)\textbf{s}^H(l)+\lambda(n) \textbf{s}(n)\textbf{s}^H(n)\right]=\sum_{l=1}^{n-1} \lambda(n)^{n-l}\textbf{r}(l)\textbf{s}^H(l)+\lambda(n)\textbf{r}(n)\textbf{s}^H(n)
\end{equation}
Then we can get
\begin{equation}
\textbf{H}(n)=\left[\sum_{l=1}^{n-1} \lambda(n)^{n-l}\textbf{r}(l)\textbf{s}^H(l)+\lambda(n)\textbf{r}(n)\textbf{s}^H(n)\right]
\left[\sum_{l=1}^{n-1} \lambda(n)^{n-l}\textbf{s}(l)\textbf{s}^H(l)+\lambda(n) \textbf{s}(n)\textbf{s}^H(n)\right]^{-1}
\end{equation}
Let:
\begin{equation}
\boldsymbol{\phi}(n)=\sum_{l=1}^{n-1} \lambda(n)^{n-l}\textbf{s}(l)\textbf{s}^H(l)+\lambda(n)\textbf{s}(n)\textbf{s}^H(n)
\end{equation}
and,
\begin{equation}
\textbf{Z}(n)=\sum_{l=1}^{n-1} \lambda(n)^{n-l}\textbf{r}(l)\textbf{s}^H(l)+\lambda(n)\textbf{r}(n)\textbf{s}^H(n)
\end{equation}
Equation (74) becomes:
\begin{equation}
\textbf{H}(n)=\textbf{Z}(n)\boldsymbol{\phi}^{-1}(n)
\end{equation}
Isolating the term corresponding to $l=n-1$ from the rest of the
summation on the right-hand side of (75), we may write:
\begin{equation}
\boldsymbol{\phi}(n)=\left[\sum_{l=1}^{n-2} \lambda(n)^{n-l}\textbf{s}(l)\textbf{s}^H(l)+\lambda(n)\textbf{s}(n-1)\textbf{s}^H(n-1)\right]\\
+\lambda(n)\textbf{s}(n)\textbf{s}^H(n)
\end{equation}
The expression inside the brackets on the right-hand
side of (78) equals $\boldsymbol{\phi}(n-1)$  assuming the
forgetting factor of the cost function is close to 1. Hence, we
have the following recursion for updating the value of
${\boldsymbol \phi}(n)$:
\begin{equation}
{\boldsymbol{\phi}}(n)=\boldsymbol{\phi}(n-1)+\lambda(n)\textbf{s}(n)\textbf{s}^H(n)
\end{equation}
Similarly, we may use (76) to derive the following recursion for
updating $\textbf{Z}(n)$:
\begin{equation}
\textbf{Z}(n)=\textbf{Z}(n-1)+\lambda(n)\textbf{r}(n)\textbf{s}^H(n)
\end{equation}
Then, using the matrix inversion lemma \cite{Haykin}, we obtain
the following recursive equation for the inverse of
$\boldsymbol{\phi}(n)$:
\begin{small}
\begin{equation}
\boldsymbol{\phi}^{-1}(n)=\boldsymbol{\phi}^{-1}(n-1)-\frac{\lambda(n)\boldsymbol{\phi}^{-1}(n-1)\textbf{s}(n)\textbf{s}^H(n)\lambda(n)\boldsymbol{\phi}^{-1}(n-1)}{1+\lambda(n)\textbf{s}^H(n)\boldsymbol{\phi}^{-1}(n-1)\textbf{s}(n)}
\end{equation}
\end{small}
For convenience of computation, let:
\begin{equation}
\textbf{P}(n)=\boldsymbol{\phi}^{-1}(n)
\end{equation}
and,
\begin{equation}
\textbf{k}(n)=\frac{\textbf{s}^H(n)\textbf{P}(n-1)}{1+\lambda(n)\textbf{s}^H(n)\textbf{P}(n-1)\textbf{s}(n)}
\end{equation}
Therefore, we may rewrite  (77) and (81) as:
\begin{equation}
\textbf{H}(n)=\textbf{Z}(n)\textbf{P}(n)
\end{equation}
\begin{equation}
\textbf{P}(n)=\textbf{P}(n-1)-\lambda(n)\textbf{P}(n-1)\textbf{s}(n)\textbf{k}(n)
\end{equation}
Then  we substitute  (80) and (85)  into (84) to obtain a
recursive equation for updating the channel matrix
$\textbf{H}(n)$:
\begin{equation}
\textbf{H}(n)=\textbf{H}(n-1)-\lambda(n)\textbf{H}(n-1)\textbf{s}(n)\textbf{k}(n)\\
+\lambda(n)\textbf{r}(n)\textbf{s}^H(n)\textbf{P}(n)
\end{equation}
By rearranging (83) , we can get:
\begin{equation}
\begin{split}
\textbf{k}(n)&=\textbf{s}^H(n)\textbf{P}(n-1)-\lambda(n)\textbf{s}^H(n)\textbf{P}(n-1)\textbf{s}(n)\textbf{k}(n)\\
&=\textbf{s}^H(n)\left[\textbf{P}(n-1)-\lambda(n)\textbf{P}(n-1)\textbf{s}(n)\textbf{k}(n)\right]\\
&=\textbf{s}^H(n)\textbf{P}(n)
\end{split}
\end{equation}
Using (87) above, we get the desired recursive equation for
updating the channel matrix $\textbf{H}(n)$:
\begin{equation}
\begin{split}
\textbf{H}(n)&=\textbf{H}(n-1)-\lambda(n)\textbf{H}(n-1)\textbf{s}(n)\textbf{k}(n)+\lambda(n)\textbf{r}(n)\textbf{k}(n)\\
&=\textbf{H}(n-1)+\lambda(n)\left[\textbf{r}(n)-\textbf{H}(n-1)\textbf{s}(n)\right]\textbf{k}(n)\\
&=\textbf{H}(n-1)+\lambda(n)\boldsymbol{\epsilon}(n)\textbf{k}(n)
\end{split}
\end{equation}
where
$\boldsymbol{\epsilon}(n)=\textbf{r}(n)-\textbf{H}(n-1)\textbf{s}(n)$
denotes the prediction error vector at time instant $n$.
//
\section*{part of the analysis of the proposed SM-NLMS channel estimation algorithm}
From (46), the update equation of the channel estimation error is:
\begin{equation}
\Delta \textbf{H}(n+1)=\Delta \textbf{H}(n)-\frac{1}{N\sigma_s^2}\textbf{e}(n)\textbf{s}^H(n)+\frac{\gamma}{N\sigma_s^2}\frac{\textbf{e}(n)}{\|\textbf{e}_0(n)\|}\textbf{s}^H(n)
\end{equation}
Let:
\begin{equation}
\textbf{A}=\Delta \textbf{H}(n)-\frac{1}{N\sigma_s^2}\textbf{e}(n)\textbf{s}^H(n)
\end{equation}
and,
\begin{equation}
\textbf{B}=\frac{\gamma}{N\sigma_s^2}\frac{\textbf{e}(n)}{\|\textbf{e}_0(n)\|}\textbf{s}^H(n)
\end{equation}
Equation (89) becomes:
\begin{equation}
\Delta \textbf{H}(n+1)=\textbf{A}+\textbf{B}
\end{equation}
From (43), we can get the output excess MSE at time instant $n+1$:
\begin{equation}
\begin{split}
J_{ex}(n+1)&=tr\{E[\textbf{s}(n+1)\textbf{s}^H(n+1)\Delta \textbf{H}^H(n+1)\Delta \textbf{H}(n+1)]\}\\
           &=tr\{E[\textbf{s}(n)\textbf{s}^H(n)\Delta \textbf{H}^H(n+1)\Delta \textbf{H}(n+1)]\}\\
           &=\psi_1+\psi_2+\psi_3+\psi_4
\end{split}
\end{equation}
Then we analyze each term separately:
\begin{equation}
\psi_1=tr\{E[\textbf{s}(n)\textbf{s}^H(n)\textbf{A}^H\textbf{A}]\}=\rho_1+\rho_2
\end{equation}
\begin{equation}
\rho_1=J_{ex}(n)-2N\sigma_s^2\frac{1}{N\sigma_s^2}J_{ex}(n)+N^2\sigma_s^4\frac{1}{N^2\sigma_s^4}J_{ex}(n)=0
\end{equation}
\begin{equation}
\rho_2=N^2\sigma_s^4M\sigma_n^2\frac{1}{N^2\sigma_s^4}=M\sigma_n^2
\end{equation}
\begin{equation}
\begin{split}
\psi_2=&tr\{E[\textbf{s}(n)\textbf{s}^H(n)\textbf{A}^H\textbf{B}]\}\\
      =&tr\{E[\textbf{s}(n)\textbf{s}^H(n)\Delta \textbf{H}^H(n)\frac{\gamma}{N\sigma_s^2}\frac{\textbf{e}(n)}{\|\textbf{e}_0(n)\|}\textbf{s}^H(n)]\}\\
       &-tr\{E[\textbf{s}(n)\textbf{s}^H(n)\frac{\gamma}{N^2\sigma_s^4}\textbf{s}(n)\frac{\textbf{e}^H(n)\textbf{e}(n)}{\|\textbf{e}_0(n)\|}\textbf{s}^H(n)]\}\\
      =&\gamma tr\{E[\textbf{s}^H(n)\Delta \textbf{H}^H(n)\frac{\textbf{e}(n)}{\|\textbf{e}_0(n)\|}]\}-\gamma E\left[\frac{\|\textbf{e}(n)\|^2}{\|\textbf{e}_0(n)\|}\right]\\
      =&\gamma tr\{E[\textbf{s}^H(n)\Delta \textbf{H}^H(n)\frac{\textbf{n}(n)+\Delta \textbf{H}(n)\textbf{s}(n)}{\|\textbf{e}_0(n)\|}]\}
       -\gamma E\left[\frac{\|\textbf{e}(n)\|^2}{\|\textbf{e}_0(n)\|}\right]\\
      =&\gamma tr\{E[\textbf{s}^H(n)\Delta \textbf{H}^H(n)\frac{\Delta \textbf{H}(n)\textbf{s}(n)}{\|\textbf{e}_0(n)\|}]\}
       -\gamma E\left[\frac{\|\textbf{e}(n)\|^2}{\|\textbf{e}_0(n)\|}\right]\\
      =&\gamma E\left[\frac{1}{\|\textbf{e}_0(n)\|}\right]J_{ex}(n)-\gamma E\left[\frac{\|\textbf{e}(n)\|^2}{\|\textbf{e}_0(n)\|}\right]
\end{split}
\end{equation}
\begin{equation}
\psi_3=tr\{E[\textbf{s}(n)\textbf{s}^H(n)\textbf{B}^H\textbf{A}]\}=\psi_2
\end{equation}
\begin{equation}
\begin{split}
\psi_4=&tr\{E[\textbf{s}(n)\textbf{s}^H(n)\textbf{B}^H\textbf{B}]\}\\
      =&tr\{E[\textbf{s}(n)\textbf{s}^H(n)\frac{\gamma^2}{N^2\sigma_s^4}\textbf{s}(n)\frac{\textbf{e}^H(n)\textbf{e}(n)}{\|\textbf{e}_0(n)\|^2}\textbf{s}^H(n)]\}\\
      =&\gamma^2 E\left[\frac{\|\textbf{e}(n)\|^2}{\|\textbf{e}_0(n)\|^2}\right]
\end{split}
\end{equation}
Finally, we can obtain the update equation of the output excess MSE:
\begin{equation}
J_{ex}(n+1)=M\sigma_n^2+2\gamma E\left[\frac{1}{\|\textbf{e}_0(n)\|}\right]J_{ex}(n)
            -2\gamma E\left[\frac{\|\textbf{e}(n)\|^2}{\|\textbf{e}_0(n)\|}\right]+\gamma^2 E\left[\frac{\|\textbf{e}(n)\|^2}{\|\textbf{e}_0(n)\|^2}\right]
\end{equation}

\end{document}